\documentclass{article}
\usepackage{spconf,amsmath,graphicx}
\usepackage{subfigure}
\usepackage{multirow}
\usepackage[urlcolor=blue]{hyperref}
\usepackage{booktabs}
\usepackage{amsfonts}
\usepackage{float}
\usepackage{amsthm,amsmath,amssymb}
\usepackage{mathrsfs}
\usepackage{bbding}
\usepackage{color}
\usepackage{amsmath}

\title{Exploiting audio-visual features with pretrained AV-HuBERT for Multi-modal Dysarthric Speech Reconstruction}
\makeatletter
\def\name#1{\gdef\@name{#1\\}}
\makeatother
\name{\em{Xueyuan Chen$^{1,2}$,Yuejiao Wang$^1$,Xixin Wu$^{1,3,*}$\thanks{* Corresponding authors. This research is supported by National Natural Science Foundation of China (62076144), CUHK Direct Grant for Research (Ref. No. 4055221), the CUHK Stanley Ho Big Data Decision Analytics Research Centre and the Centre for Perceptual and Interactive Intelligence.},Disong Wang$^{3}$,Zhiyong Wu$^{1,2,*}$,Xunying Liu$^{1}$,Helen Meng$^{1,2,3}$}}

\address{
    $^1$ Department of Systems Engineering and Engineering Management, \\
         The Chinese University of Hong Kong, Hong Kong SAR, China\\
    $^2$ Tsinghua-CUHK Joint Research Center for Media Sciences, Technologies and Systems, \\
    Shenzhen International Graduate School, Tsinghua University, Shenzhen, China\\
    $^3$ Vocal Engineering Technologies Limited, Hong Kong SAR, China\\
    \small{ 
        \{xychen, wangy, wuxx, dswang, zywu, xyliu, hmmeng\}@se.cuhk.edu.hk, 
    }}

\begin{document}
\ninept
\maketitle
\begin{abstract}
\vspace{-3pt}
Dysarthric speech reconstruction (DSR) aims to transform dysarthric speech into normal speech by improving the intelligibility and naturalness.
This is a challenging task especially for patients with severe dysarthria and speaking in complex, noisy acoustic environments.
To address these challenges, we propose a novel multi-modal framework to utilize visual information, e.g., lip movements, in DSR as extra clues for reconstructing the highly abnormal pronunciations. The multi-modal framework consists of: (i) a multi-modal encoder to extract robust phoneme embeddings from dysarthric speech with auxiliary visual features; (ii) a variance adaptor to infer the normal phoneme duration and pitch contour from the extracted phoneme embeddings; (iii) a speaker encoder to encode the speaker’s voice characteristics; and (iv) a mel-decoder to generate the reconstructed mel-spectrogram based on the extracted phoneme embeddings, prosodic features and speaker embeddings. Both objective and subjective evaluations conducted on the commonly used UASpeech corpus show that our proposed approach can achieve significant improvements over baseline systems in terms of speech intelligibility and naturalness, especially for the speakers with more severe symptoms. Compared with original dysarthric speech, the reconstructed speech achieves 42.1\% absolute word error rate reduction for patients with more severe dysarthria levels.\footnote[1]{\href{https://Chenxuey20.github.io/MMDSR}{Audio samples: https://Chenxuey20.github.io/MMDSR}}

\end{abstract}


\begin{figure*}[ht]
\centering
\includegraphics[width=1.88\columnwidth]{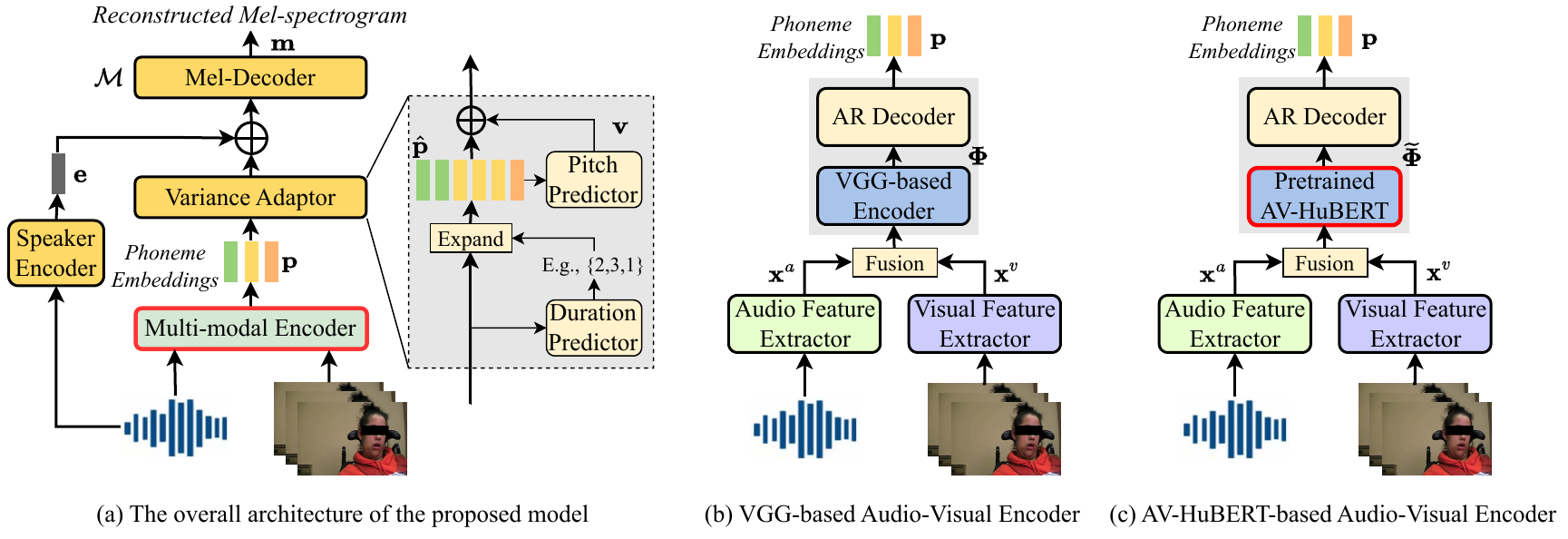}
\vspace{-12pt}
\caption{Diagram of the proposed multi-modal DSR system, where (a) illustrates the overall architecture with Multi-modal Encoder, Variance Adaptor, Speaker Encoder and Mel-Decoder, (b) and (c) show two different Multi-modal Encoders to be compared respectively.}
\vspace{-10pt}
\label{fig:model_structure}
\end{figure*}

\vspace{-10pt}
\section{introduction}
\vspace{-7pt}

Dysarthria is a common form of speech disorders associated
with neuromotor conditions \cite{darley1975motor}, such as Parkinson's disease and
cerebral palsy, as well as brain damages due to stroke or head injuries \cite{whitehill2000speech}. 
Dysarthria leads to severe degradation of speech quality, highly variable voice characteristics and large deviation from normal speech, significantly affecting the communication between dysarthric patients and their family or caregivers \cite{kain2007improving}. 
Dysarthric speech reconstruction (DSR) is among the effective solutions to assist the communication by converting dysarthric speech to normal speech with higher intelligibility and naturalness.

DSR is a challenging task that attracts many research efforts to improve the speech quality, intelligibility and naturalness of reconstructed speech.
The voice banking-based method employs the speech recordings of dysarthric patients that are collected before their pronunciation ability deteriorates to build personalized text-to-speech (TTS) synthesis systems  \cite{yamagishi2012speech}.
However, this approach is constrained by the condition of having the speech recordings before dysarthria and not sutiable for all patients \cite{chen2022hilvoice}.
Some studies try to tackle the reconstruction problem by voice conversion (VC) techniques 
which adjust dysarthric speech signals to be more intelligible and natural while keep the content unchanged.
Among them, the rule-based VC modifies the temporal or frequency characteristics of speech according to specifically designed rules
\cite{kumar2016improving}.
The statistical VC approach creates a mapping function between the acoustic features of dysarthric and those of normal speech,
based on Gaussian mixture model (GMM) \cite{kain2007improving}, non-negative matrix factorization (NMF) \cite{fu2016joint}, and partial least square (PLS) \cite{aihara2017phoneme}, etc.
After that, an end-to-end VC (E2E-VC)-based DSR system is proposed with cross-modal knowledge distillation (KD) 
by distilling a speech encoder from a pretrained speech recognition model to replace the
text encoder of a sequence-to-sequence (seq2seq) TTS system \cite{wang2020end}.
Compared with the naive cascaded system that feeds the automatic speech recognition (ASR) results to a TTS model,
E2E-VC does not constrain the intermediate representations to text characters and can generate speech with lower error and higher fidelity.
Furthermore, an additional prosody corrector and 
deep speaker encoder are added to improve the prosody and speaker similarity \cite{wang2020learning}.

Previous works mainly focus on the dysarthric patients with lower severity levels \cite{wang2020end,wang2020learning}.
However, the intelligibility and naturalness of reconstructed speech are still unsatisfactory especially for the highly serious dysarthric speakers.
Besides, the complex noisy environments also have a 
critical
effect on the semantic representation extraction.
In these cases, it is insufficient to rely solely on the information from the audio modality.
In fact, 
human perception of speech is intrinsically multi-modal, involving audition and vision.
Previous researches have shown that incorporating visual modality can improve the performance of ASR systems
in noisy environments 
\cite{zhang2019robust}
and on disordered speech 
\cite{hu2019cuhk}.

With all listed imperfections taken into consideration, this paper proposes a multi-modal framework that introduces visual features for improved DSR performance based on our previous work \cite{wang2020learning}.
Firstly, a multi-modal encoder is specially designed to extract more robust phoneme embeddings from dysarthric speech and visual inputs.
Secondly, we use a variance adaptor to infer the normal phoneme duration and pitch values from the extracted phoneme embeddings.
Thirdly, the mel-decoder takes robust phoneme embeddings and  normal prosody features
as inputs to generate the converted speech, conditioned on the speaker embedding that is learned via a well-trained speaker encoder.
The main contributions of this paper include:
\begin{itemize}
\item To the best of our knowledge, this is the first time to propose a multi-modal DSR model
which uses both the audio and visual features to reconstruct the dysarthric speech.

\item Two different multi-modal encoders are proposed and compared with the audio-only encoder.
Specially, we adopt the pretrained AV-HuBERT to this task to extract more accurate semantic representations.

\item Both objective and subjective experimental results show that our proposed multi-modal DSR system achieves significant improvements in terms of speech intelligibility and naturalness,
especially for the speakers with more severe symptoms.
\end{itemize}

\vspace{-8pt}
\section{methodology}
\vspace{-5pt}
Our proposed  multi-modal DSR model is illustrated in Fig. \ref{fig:model_structure} (a).
It mainly consists of a
multi-modal encoder, a variance adaptor, a speaker encoder and a mel-decoder.
Specifically, the multi-modal encoder strives to extract robust phoneme embeddings from 
dysarthric audio and video inputs.
The variance adaptor is introduced to explicitly model the 
prosodic features.
The speaker encoder is used to learn the speaker embedding.
The mel-decoder takes phoneme embeddings and prosody features as inputs to generate the converted speech, conditioned on the speaker embedding.

\vspace{-5pt}
\subsection{Multi-modal Encoder for Phoneme Embeddings Extraction}
To 
reconstruct the linguistic content of original dysarthric speech, a multi-modal encoder is used to extract robust linguistic representations.
Following \cite{wang2020learning}, we adopt the multi-modal encoder outputs, i.e., the phoneme probability distribution, as the phoneme embeddings, which are denoted as $\mathbf{p}$.
In order to verify the effect of visual information in the DSR task, 
two different multi-modal encoders are proposed and compared with the audio-only encoder as follows.

\vspace{-5pt}
\subsubsection{Audio-only Encoder}
\label{sec:audio speech encoder}
\begin{figure}[ht]
\centering
\vspace{-5pt}
\includegraphics[width=0.67\columnwidth]{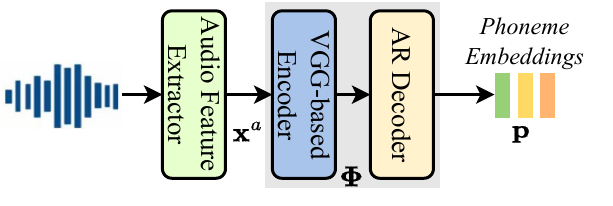}
\vspace{-10pt}
\caption{Audio-only Encoder}
\vspace{-10pt}
\label{fig:audio_encoder}
\end{figure}

As shown in Fig. \ref{fig:audio_encoder}, the audio-only encoder has a similar architecture as in \cite{wang2020learning},
which contains an audio feature extractor and a common auto-regressive (AR) seq2seq ASR model $\mathbf{\Phi}$.
As the dysarthric audio has strong background noise which degrades the speech encoder performance,
we first adopt log-MMSE speech enhancement algorithm \cite{ephraim1985speech} to preprocess the audio, and then 80-dimension filter banks (FBKs)+$\Delta$ features are extracted as the audio features $\mathbf{x}^a$.
Upon the audio features, a seq2seq ASR model $\mathbf{\Phi}$ consisting of a VGG-based encoder \cite{simonyan2014very} and an AR decoder with connectionist temporal classification (CTC) is adopted. The encoder is composed of a 6-layer VGG extractor and 5 bidirectional long short-term memory (BLSTM) layers with 512 units per direction. The decoder contains a 512-dimensional location-aware attention module and 2 LSTM layers with 1024 units per layer.
Finally, the ASR model $\mathbf{\Phi}$ outputs phoneme embeddings $\mathbf{p}$, which can be represented as $\mathbf{p}=\mathbf{\Phi} (\mathbf{x}^a)$.


\vspace{-5pt}
\subsubsection{VGG-based Audio-Visual Encoder}
\label{sec:audio-visual speech encoder}
Inspired by the success of audio-visual speech recognition \cite{afouras2018deep, liu2019exploiting},
we develop a VGG-based audio-visual encoder to introduce the visual feature inputs based on the audio-only encoder.
As illustrated in Fig. \ref{fig:model_structure} (b), 
the visual features $\mathbf{x}^v$ are extracted by a visual feature extractor.
Firstly, we force-align each video segment with the corresponding audio features by upsampling.
After that, an off-the-shelf face alignment network \cite{bulat2017far} is employed to detect lip landmarks on the upsampled video data,
followed by affine transformation to make the detected lip regions horizontal.
Finally, discrete cosine transform (DCT) and linear discriminant analysis (LDA) are used to downsize and obtain the visual features $\mathbf{x}^v$.
Since the extracted visual features $\mathbf{x}^v$ are temporally aligned with the audio features $\mathbf{x}^a$, we further take the common operation of concatenating both audio and visual features along the feature dimension followed by a fully-connected (FC) layer to fuse and get the audio-visual features, which are further fed into the following seq2seq ASR model $\mathbf{\Phi}$.
It can be described as 
$\mathbf{p}=\mathbf{\Phi} (\mathbf{W}_1 (\mathbf{x}^a\oplus\mathbf{x}^v)+\mathbf{z}_1)$,
where $\oplus$
is concatenation along the feature dimension, $\mathbf{W}_1$ and $\mathbf{z}_1$ are FC-layer parameters.

\vspace{-5pt}
\subsubsection{AVHuBERT-based Audio-Visual Encoder}
\label{sec:av-hubert speech encoder}
AV-HuBERT \cite{shi2022avhubert} is a self-supervised learning framework for audio-visual speech representations by masking multi-stream video inputs and predicting multi-modal hidden units that are automatically discovered and iteratively refined.
AV-HuBERT can learn powerful audio-visual speech representations benefiting both lip-reading and ASR.
Therefore, as shown in Fig. \ref{fig:model_structure} (c), we adopt the well-pretrained AV-HuBERT transformer encoder to replace the VGG-based encoder in the multi-modal encoder.
Similarly, the audio-visual feature fusion before the backbone transformer encoder is also concatenation along the feature dimension,
which is consistent with the original AV-HuBERT design. The fused features are fed to a FC layer to map the extracted features to the corresponding dimension in the AV-HuBERT.
Likewise, the process is represented as
$\mathbf{p}=\mathbf{\widetilde{\Phi}} (\mathbf{W}_2 (\mathbf{x}^a\oplus\mathbf{x}^v)+\mathbf{z}_2)$,
where $\mathbf{\widetilde{\Phi}}$ is the improved ASR model with AV-HuBERT, $\mathbf{W}_2$ and $\mathbf{z}_2$ are the FC-layer parameters.

\vspace{-5pt}
\subsection{Variance Adaptor for Explicit Prosody Modeling}
Inspired by the prosody modeling in TTS systems \cite{chen2022character,chen2022unsupervised,chen2023stylespeech}, 
we adopt a variance adaptor to reconstruct the dysarthric prosody to its normal version by explicit prosody modeling.
As shown in Fig. \ref{fig:model_structure} (a), the variance adaptor contains a phoneme duration predictor and a pitch ($F_0$) predictor.
Both duration and $F_0$ predictors adopt the same structure, which consists of 3 bidirectional gated recurrent unit (BiGRU) layers with 256 units per direction, 3 convolution layers with kernel sizes of 5, 9 and 19 respectively, and a 1-dimensional FC layer.
Both duration and $F_0$ predictors are trained with $L1$ loss using normal speech:
(1) For duration prediction, the inputs are phoneme embeddings $\mathbf{p}$ extracted by the 
multi-modal encoder using the teacher-forcing mode.
The targets are ground-truth phoneme durations.
(2) For $F_0$ prediction, the expanded phoneme embeddings $\hat{\mathbf{p}}$ based on the ground-truth phoneme durations are used as the inputs, 
and the targets are the ground-truth $F_0$, denoted by $\mathbf{v}$.
When the duration and $F_0$ predictors are well-trained,
the variance adaptor is expected to infer normal 
prosody values to replace their abnormal counterparts for speech reconstruction.

\vspace{-5pt}
\subsection{Speaker Encoder and Mel-Decoder }
To preserve the original speaker timbre,
a scalable and accurate neural network for speaker verification task is adopted as the speaker encoder to produce a fixed-dimensional deep speaker embedding (DSE) $\mathbf{e}$
for a speech utterance, following \cite{wan2018generalized}. 
The speaker encoder is trained using a generalized end-to-end speaker verification loss, 
so that the DSEs extracted from one speaker's utterances have high similarity, and those from different speakers have low similarity.

The mel-decoder denoted as $\mathcal{M}$ is finally used to generate the reconstructed mel-spectrogram $\mathbf{m}$,
based on the extracted content representation, prosodic features and speaker embeddings.
Specifically, the expanded phoneme embeddings $\hat{\mathbf{p}}$, the pitch value $\mathbf{v}$ and the repeated DSE $\hat{\mathbf{e}}$ are first concatenated for each time step and then fed into the mel-decoder,
which can be represented as
$\mathbf{m} = \mathcal{M}(\hat{\mathbf{p}} \oplus \mathbf{v} \oplus \hat{\mathbf{e}})$.
The mel-decoder is trained with normal speech data to minimize a loss $L$ (e.g., L1 loss) that measures the distance between the predicted and ground-truth mel-spectrograms.

\vspace{-5pt}
\section{Experiments}

\begin{table*}[!htb]
\caption{Results of multi-modal encoder comparison with audio-only encoder and objective speech intelligibility comparison, where `$\Delta1$' denotes the WER or PER improvement between the best and worst DSR systems,  `$\Delta2$' denotes the WER improvement between the best DSR system and original dysarthric speech.}
\label{table:results of per and wer}
\centering
\resizebox{1.9\columnwidth}{!}{
\begin{tabular}{cccccccccc}
            \toprule
\multirow{2}{*}{\textbf{Speaker}}     & \textbf{Speech}          & \multirow{2}{*}{\textbf{Metrics}} & \multirow{2}{*}{\textbf{Original}} & \textbf{Original}  & \multirow{2}{*}{\textbf{A-DSR}}   & \multirow{2}{*}{\textbf{AV-DSR}}   & \textbf{AVHuBERT}   & \textbf{$\Delta 1$:} & \textbf{$\Delta 2$:}\\
            & \textbf{Intelligibility} &         &          & \textbf{(Mel+PWG)} &   &   &\textbf{-DSR}   & \textbf{Best-Worst} & \textbf{Best-Original} \\
          \midrule
\multirow{2}{*}{\textbf{M12}}& \multirow{2}{*}{Very Low (7.4\%)} & PER & -       & -       & 57.0\% & 53.0\% & \textbf{47.0\%} & \textbf{-10.0\%} & - \\
            &              & WER & 98.0\% & 99.0\% & 65.7\% & 60.6\% & \textbf{55.9\%} & \textbf{-9.8\%}  & \textbf{-42.1\%} \\
\multirow{2}{*}{\textbf{F02}} & \multirow{2}{*}{Low (29\%)}  & PER & -       & -       & 56.0\% & 52.0\% & \textbf{48.0\%} & -8.0\%  & -  \\
             &             & WER & 93.4\%  & 97.8\%  & 62.5\% & 59.2\% & \textbf{56.1\%} & -6.4\%  & -37.3\%  \\
\multirow{2}{*}{\textbf{M16}} & \multirow{2}{*}{Low (43\%)}  & PER & -       & -       & 48.0\% & 46.0\% & \textbf{39.0\%} & -9.0\%  & -  \\
             &             & WER & 81.6\% & 90.1\% & 60.3\% & 57.0\% & \textbf{51.3\%} & -9.0\%  & -30.3\%  \\
\multirow{2}{*}{\textbf{F04}} & \multirow{2}{*}{Middle (62\%)}  & PER & -       & -       & 43.0\% & 42.0\% & \textbf{36.0\%} & -7.0\%  & -  \\
             &             & WER & 67.6\% & 81.3\% & 55.6\% & 55.4\% & \textbf{48.0\%} & -7.6\%  & -19.6\%  \\
             \midrule
\multirow{2}{*}{\textbf{Average}} & \multirow{2}{*}{-}       & PER & -       & -       & 51.0\% & 48.3\% & \textbf{42.5\%} & -8.5\%  & -  \\
             &             & WER & 85.2\% & 92.1\% & 61.0\% & 58.1\% & \textbf{52.8\%} & -8.2\%  & -32.4\%  \\
        \bottomrule
\end{tabular}
}
\end{table*}

\vspace{-5pt}
\subsection{Experimental Settings}
Experiments are conducted on the UASpeech \cite{kim2008dysarthric}, LJSpeech \cite{ito2017lj} and VCTK \cite{veaux2016superseded} datasets.
The UASpeech corpus is a benchmark disordered speech corpus, 
which is recorded by an 8-channel microphone array and a video camera with some background noise. 7 channels are segmented into single word audios. 
We use parallel WaveGAN (PWG) \cite{yamamoto2020parallel} as vocoder to synthesize the waveform from the converted mel-spectrograms.
We use the VCTK Corpus with 105 native speakers to train the speaker encoder, mel-decoder and PWG. 
The LJSpeech with normal speech from a female speaker is used to train the duration predictor and $F_0$ predictor in the variance adaptor.

\begin{figure}[t]
\centering	\includegraphics[width=0.8\columnwidth]{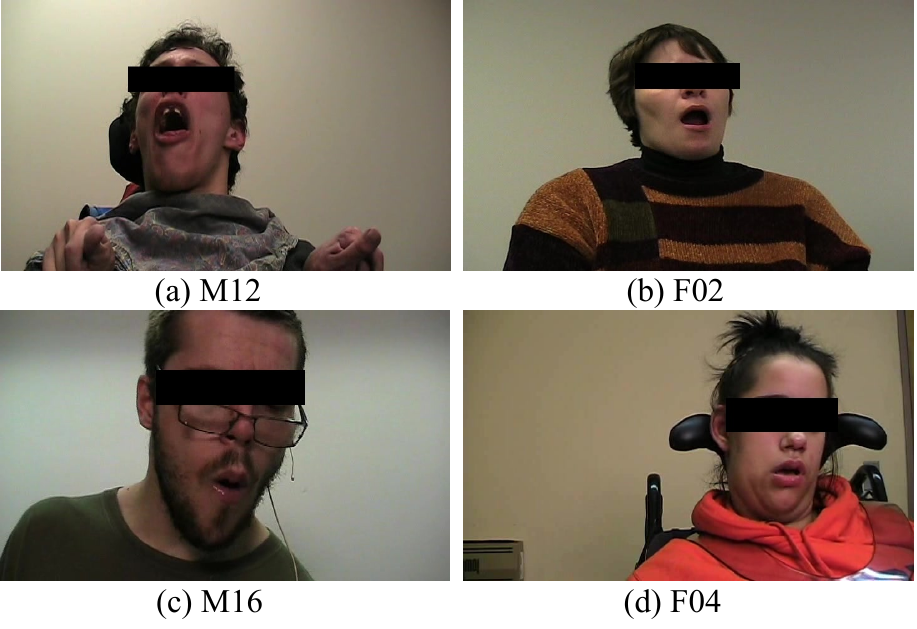}
\vspace{-8pt}
\caption{Example video snapshots of four UASpeech speakers M12, F02, M16 and F04 with various angles facing the camera.}
\label{fig:video_snapshots}
\vspace{-10pt}
\end{figure}

As the dysarthria severity is varied among different patients,
which increases the modeling difficulties to build a generalized DSR system for all patients,
four speaker-dependent DSR systems are separately built for the four selected speakers (M12, F02, M16 and F04) with the lowest speech intelligibility.
The example video snapshots and speech intelligibility (the correct percentage in word transcription tasks) of the selected four speakers are shown in Fig. \ref{fig:video_snapshots} and Table \ref{table:results of per and wer} respectively.
To verify the effectiveness of visual information for the DSR task,
three types of model settings 
are compared:

\begin{itemize}
\item \textbf{A-DSR}: The audio-only encoder (described in \ref{sec:audio speech encoder}) is used for the DSR model.

\item \textbf{AV-DSR}: The VGG-based audio-visual encoder (described in \ref{sec:audio-visual speech encoder}) is used for the DSR model.

\item \textbf{AVHuBERT-DSR}: The AV-HuBERT-based audio-visual encoder (described in \ref{sec:av-hubert speech encoder}) is used for the DSR model,
where the open-source pretrained `AV-HuBERT Base' model\footnote[1]{\href{https://github.com/facebookresearch/av_hubert}{https://github.com/facebookresearch/av\_hubert}} is adopted in our experiments.

\end{itemize}

The multi-modal encoders and audio-only encoder are first trained on the whole dysarthric speech dataset of all speakers, and then finetuned on the target speaker to improve phoneme prediction accuracy.
Adadelta optimizer \cite{zeiler2012adadelta} with learning rate of 1 and batch size 8 is applied for the training and finetuning 
for 1M and 2k steps, respectively. 
Both duration and $F_0$ predictors are trained by the Adam optimizer \cite{kingma2014adam} with learning rate of 0.001 and batch size of 16 for 30k steps. 
The parameter settings and training details of speaker encoder and mel-decoder are the same as \cite{liu2020end}.

\vspace{-5pt}
\subsection{Experimental Results and Analysis}

\subsubsection{Multi-modal Encoder Comparison with Audio-only Encoder}
In order to verify the effectiveness of visual features,
we first decode the phoneme sequences from the phoneme embeddings output from the two different multi-modal encoders and the audio-only encoder, respectively.
Phoneme error rate (PER) is calculated with the ground-truth phoneme sequence for the three decoded phoneme sequences.
The results are shown in Table \ref{table:results of per and wer}.

In general, the PERs for all speakers are still at a high level, as the DSR task is very challenging. Note that in our experiments, we specially select the speakers with lowest speech intelligibility. Hence, the obtained PER results indicate the effectiveness of the three encoders.
After adding visual features, the performance of AV-DSR is improved compared to A-DSR,
and the pretrained AV-HuBERT model achieves the best results,
with an average PER reduction of 8.5\%.
It is worth noting that as the speech intelligibility decreases,
i.e., the difficulty of reconstruction gradually increases,
the improvement brought by visual features becomes more obvious,
with the PER reduction increasing from 7.0\% to 10.0\%.

\vspace{-5pt}
\subsubsection{Objective Speech Intelligibility Comparison}
To show the content intelligibility improvement of final reconstructed speech compared with the `\textbf{Original}' dysarthric speech,
a publicly released ASR model, Whisper \cite{radford2023robust}, is used to obtain the word error rate (WER) with greedy decoding.
We also report the result for copy synthesis `\textbf{Original (Mel+PWG)}' that uses the original mel-spectrograms to synthesize the waveform by using PWG vocoder.

As can be observed in Table Table \ref{table:results of per and wer},
`Original (Mel+PWG)' is inferior to `Original' for all four speakers, which indicates that the PWG vocoder tends to degrade the speech quality.
Similar to the PER results of multi-modal and audio-only encoders, the performance of AV-DSR with visual features is generally improved compared with A-DSR,
and the pretrained AV-HuBERT model further improves the results for all the 4 speakers,
with an average WER reduction of 8.2\%.
Compared with the original dysarthric speech, the reconstructed speech of AVHuBERT-DSR achieves an average WER reduction of 32.3\% for all the speakers and the best WER reduction of 42.1\% for the most severe speaker M12. 
Similarly and importantly, as the original speech intelligibility decreases, the WER reduction, either `Best-Worst' between the best DSR system (AVHuBERT-DSR) and the worst DSR system (A-DSR) or `Best-Original' between the best DSR system (AVHuBERT-DSR) and the original dysarthric speech,
gradually becomes more obvious.
Both the PER and WER results illustrate that it is not enough to rely solely on audio features for the highly severe patients,
and visual features can play an important role in these cases.

\begin{figure}[htb]
\centering	\includegraphics[width=0.8\columnwidth]{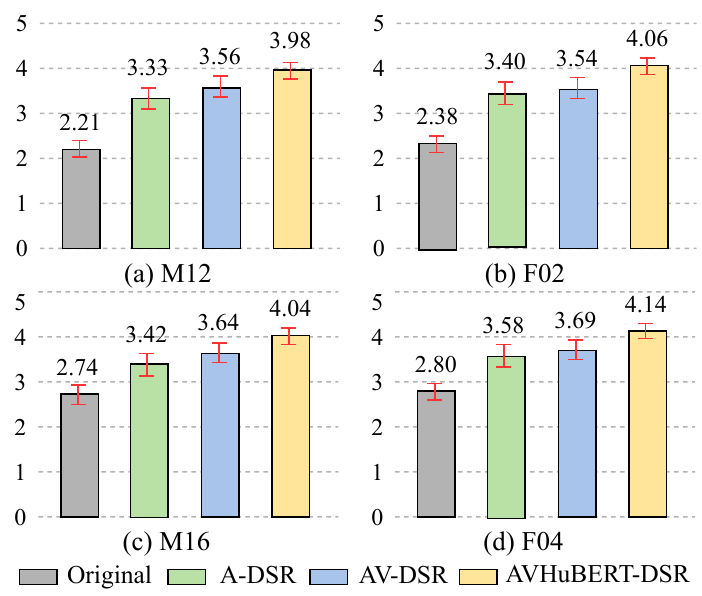}
\vspace{-10pt}
\caption{Comparison results of MOS with 95\% confidence in terms of speech intelligibility and naturalness.}
\vspace{-10pt}
\label{fig:MOS}
\end{figure}

\vspace{-5pt}
\subsubsection{Subjective Speech Naturalness Comparison}
Mean opinion score (MOS) tests are conducted to verify the effectiveness of the proposed method to improve the intelligibility and naturalness of reconstructed speech.
10 subjects are invited to 5-scale mean opinion score (MOS) tests (1-bad, 2-poor, 3-fair, 4-good, 5-excellent) to evaluate speech intelligibility and naturalness.
We random select 10 testing utterances from speaker M12, F02, M16 and F04 respectively.
Fig. \ref{fig:MOS} shows the MOS test results for the speakers,
where `Original' denotes the original dysarthric speech.

On one hand, compared with the original speech, we can observe that all the three DSR systems achieve significant improvements and AVHuBERT-DSR achieves the best result with a MOS score around 4.
On the other hand, compared with the A-DSR system that only takes audio features as inputs,
AV-DSR and AVHuBERT show better MOS scores for all the speakers.
Specifically, the improvement brought by introducing visual features is the most obvious for the most severe patient M12, changing from 3.33 of A-DSR to 3.98 of AVHuBERT-DSR.
The MOS results are consistent with the objective experimental results, which further demonstrate that visual features are very helpful for the DSR task, especially for the patients with more severe dysarthria levels.

\vspace{-9pt}
\section{Conclusion}
\vspace{-7pt}
This paper proposes to incorporate visual features for improving dysarthric speech reconstruction (DSR).
Two different multi-modal encoders have been developed and compared with the audio-only encoder.
Specifically, we adopt the pretrained AV-HuBERT to this task to extract more accurate semantic representations.
Both objective and subjective experimental results show that the proposed approach can achieve significant improvements in terms of speech intelligibility and naturalness, especially for the speakers with
more severe symptoms.

\bibliographystyle{IEEEbib}
\bibliography{refs}

\begin{thebibliography}{10}

\bibitem{darley1975motor}
Frederic~L Darley, Arnold~Elvin Aronson, and Joe~Robert Brown,
\newblock ``Motor speech disorders,''
\newblock {\em (No Title)}, 1975.

\bibitem{whitehill2000speech}
Tara~L Whitehill and Valter Ciocca,
\newblock ``Speech errors in cantonese speaking adults with cerebral palsy,''
\newblock {\em Clinical linguistics \& phonetics}, vol. 14, no. 2, pp.
  111--130, 2000.

\bibitem{kain2007improving}
Alexander~B Kain, John-Paul Hosom, Xiaochuan Niu, Jan~PH Van~Santen, Melanie
  Fried-Oken, and Janice Staehely,
\newblock ``Improving the intelligibility of dysarthric speech,''
\newblock {\em Speech communication}, vol. 49, no. 9, pp. 743--759, 2007.

\bibitem{yamagishi2012speech}
Junichi Yamagishi, Christophe Veaux, Simon King, and Steve Renals,
\newblock ``Speech synthesis technologies for individuals with vocal
  disabilities: Voice banking and reconstruction,''
\newblock {\em Acoustical Science and Technology}, vol. 33, no. 1, pp. 1--5,
  2012.

\bibitem{chen2022hilvoice}
Xueyuan Chen, Qiaochu Huang, Xixin Wu, Zhiyong Wu, and Helen Meng,
\newblock ``Hilvoice: Human-in-the-loop style selection for elder-facing speech
  synthesis,''
\newblock in {\em ISCSLP}. IEEE, 2022, pp. 86--90.

\bibitem{kumar2016improving}
S~Arun Kumar and C~Santhosh Kumar,
\newblock ``Improving the intelligibility of dysarthric speech towards
  enhancing the effectiveness of speech therapy,''
\newblock in {\em 2016 International Conference on Advances in Computing,
  Communications and Informatics (ICACCI)}. IEEE, 2016, pp. 1000--1005.

\bibitem{fu2016joint}
Szu-Wei Fu, Pei-Chun Li, Ying-Hui Lai, Cheng-Chien Yang, Li-Chun Hsieh, and
  Yu~Tsao,
\newblock ``Joint dictionary learning-based non-negative matrix factorization
  for voice conversion to improve speech intelligibility after oral surgery,''
\newblock {\em IEEE Transactions on Biomedical Engineering}, vol. 64, no. 11,
  pp. 2584--2594, 2016.

\bibitem{aihara2017phoneme}
Ryo Aihara, Tetsuya Takiguchi, and Yasuo Ariki,
\newblock ``Phoneme-discriminative features for dysarthric speech
  conversion.,''
\newblock in {\em Interspeech}, 2017, pp. 3374--3378.

\bibitem{wang2020end}
Disong Wang, Jianwei Yu, Xixin Wu, Songxiang Liu, Lifa Sun, Xunying Liu, and
  Helen Meng,
\newblock ``End-to-end voice conversion via cross-modal knowledge distillation
  for dysarthric speech reconstruction,''
\newblock in {\em ICASSP 2020}. IEEE, 2020, pp. 7744--7748.

\bibitem{wang2020learning}
Disong Wang, Songxiang Liu, Lifa Sun, Xixin Wu, Xunying Liu, and Helen Meng,
\newblock ``Learning explicit prosody models and deep speaker embeddings for
  atypical voice conversion,''
\newblock {\em arXiv preprint arXiv:2011.01678}, 2020.

\bibitem{zhang2019robust}
Shiliang Zhang, Ming Lei, Bin Ma, and Lei Xie,
\newblock ``Robust audio-visual speech recognition using bimodal dfsmn with
  multi-condition training and dropout regularization,''
\newblock in {\em ICASSP 2019}. IEEE, 2019, pp. 6570--6574.

\bibitem{hu2019cuhk}
Shoukang Hu, Shansong Liu, Heng~Fai Chang, Mengzhe Geng, Jiani Chen, Lau~Wing
  Chung, To~Ka Hei, Jianwei Yu, Ka~Ho Wong, Xunying Liu, et~al.,
\newblock ``The cuhk dysarthric speech recognition systems for english and
  cantonese.,''
\newblock in {\em INTERSPEECH}, 2019, pp. 3669--3670.

\bibitem{ephraim1985speech}
Yariv Ephraim and David Malah,
\newblock ``Speech enhancement using a minimum mean-square error log-spectral
  amplitude estimator,''
\newblock {\em IEEE transactions on acoustics, speech, and signal processing},
  vol. 33, no. 2, pp. 443--445, 1985.

\bibitem{simonyan2014very}
Karen Simonyan and Andrew Zisserman,
\newblock ``Very deep convolutional networks for large-scale image
  recognition,''
\newblock {\em arXiv preprint arXiv:1409.1556}, 2014.

\bibitem{afouras2018deep}
Triantafyllos Afouras, Joon~Son Chung, Andrew Senior, Oriol Vinyals, and Andrew
  Zisserman,
\newblock ``Deep audio-visual speech recognition,''
\newblock {\em IEEE transactions on pattern analysis and machine intelligence},
  vol. 44, no. 12, pp. 8717--8727, 2018.

\bibitem{liu2019exploiting}
Shansong Liu, Shoukang Hu, Yi~Wang, Jianwei Yu, Rongfeng Su, Xunying Liu, and
  Helen Meng,
\newblock ``Exploiting visual features using bayesian gated neural networks for
  disordered speech recognition.,''
\newblock in {\em INTERSPEECH}, 2019, pp. 4120--4124.

\bibitem{bulat2017far}
Adrian Bulat and Georgios Tzimiropoulos,
\newblock ``How far are we from solving the 2d \& 3d face alignment
  problem?(and a dataset of 230,000 3d facial landmarks),''
\newblock in {\em ICCV}, 2017, pp. 1021--1030.

\bibitem{shi2022avhubert}
Bowen Shi, Wei-Ning Hsu, Kushal Lakhotia, and Abdelrahman Mohamed,
\newblock ``Learning audio-visual speech representation by masked multimodal
  cluster prediction,''
\newblock {\em arXiv preprint arXiv:2201.02184}, 2022.

\bibitem{chen2022character}
Xueyuan Chen, Changhe Song, Yixuan Zhou, Zhiyong Wu, Changbin Chen, Zhongqin
  Wu, and Helen Meng,
\newblock ``A character-level span-based model for mandarin prosodic structure
  prediction,''
\newblock in {\em ICASSP 2022}. IEEE, 2022, pp. 7602--7606.

\bibitem{chen2022unsupervised}
Xueyuan Chen, Shun Lei, Zhiyong Wu, Dong Xu, Weifeng Zhao, and Helen Meng,
\newblock ``Unsupervised multi-scale expressive speaking style modeling with
  hierarchical context information for audiobook speech synthesis,''
\newblock in {\em COLING}, 2022, pp. 7193--7202.

\bibitem{chen2023stylespeech}
Xueyuan Chen, Xi~Wang, Shaofei Zhang, Lei He, Zhiyong Wu, Xixin Wu, and Helen
  Meng,
\newblock ``Stylespeech: Self-supervised style enhancing with vq-vae-based
  pre-training for expressive audiobook speech synthesis,''
\newblock {\em arXiv preprint arXiv:2312.12181}, 2023.

\bibitem{wan2018generalized}
Li~Wan, Quan Wang, Alan Papir, and Ignacio~Lopez Moreno,
\newblock ``Generalized end-to-end loss for speaker verification,''
\newblock in {\em ICASSP 2018}. IEEE, 2018, pp. 4879--4883.

\bibitem{kim2008dysarthric}
Heejin Kim, Mark Hasegawa-Johnson, Adrienne Perlman, Jon Gunderson, Thomas~S
  Huang, Kenneth Watkin, and Simone Frame,
\newblock ``Dysarthric speech database for universal access research,''
\newblock in {\em Ninth Annual Conference of the International Speech
  Communication Association}, 2008.

\bibitem{ito2017lj}
Keith Ito and Linda Johnson,
\newblock ``The lj speech dataset,''
\newblock 2017.

\bibitem{veaux2016superseded}
Christophe Veaux, Junichi Yamagishi, Kirsten MacDonald, et~al.,
\newblock ``Superseded-cstr vctk corpus: English multi-speaker corpus for cstr
  voice cloning toolkit,''
\newblock 2016.

\bibitem{yamamoto2020parallel}
Ryuichi Yamamoto, Eunwoo Song, and Jae-Min Kim,
\newblock ``Parallel wavegan: A fast waveform generation model based on
  generative adversarial networks with multi-resolution spectrogram,''
\newblock in {\em ICASSP 2020}. IEEE, 2020, pp. 6199--6203.

\bibitem{zeiler2012adadelta}
Matthew~D Zeiler,
\newblock ``Adadelta: an adaptive learning rate method,''
\newblock {\em arXiv preprint arXiv:1212.5701}, 2012.

\bibitem{kingma2014adam}
Diederik~P Kingma and Jimmy Ba,
\newblock ``Adam: A method for stochastic optimization,''
\newblock {\em arXiv preprint arXiv:1412.6980}, 2014.

\bibitem{liu2020end}
Songxiang Liu, Disong Wang, Yuewen Cao, Lifa Sun, Xixin Wu, Shiyin Kang,
  Zhiyong Wu, Xunying Liu, Dan Su, Dong Yu, et~al.,
\newblock ``End-to-end accent conversion without using native utterances,''
\newblock in {\em ICASSP 2020}. IEEE, 2020, pp. 6289--6293.

\bibitem{radford2023robust}
Alec Radford, Jong~Wook Kim, Tao Xu, Greg Brockman, Christine McLeavey, and
  Ilya Sutskever,
\newblock ``Robust speech recognition via large-scale weak supervision,''
\newblock in {\em ICML}. PMLR, 2023, pp. 28492--28518.

\end{thebibliography}
\end{document}